\documentclass[aps,prl,twocolumn,superscriptaddress,floatfix,citeautoscript,preprintnumbers]{revtex4-1}
\usepackage{graphicx,color}
\usepackage{amssymb}
\usepackage{amsmath}
\usepackage{epstopdf}
\usepackage[normalem]{ulem}
\usepackage{comment}
\usepackage{breakcites}
\usepackage[breaklinks=true]{hyperref}
\usepackage[caption=false]{subfig}
\usepackage{enumitem}
\usepackage{soul}
\usepackage{bbold,dsfont}

\def\Id{\mathbb{1}}
\def\His{H_{\mathrm{I}}}

\def\Z2{Z_{\mathrm{st2}}}

\newcommand{\tlcc}{TLCC}

\newcommand{\beq}{\begin{equation}}
\newcommand{\eeq}{\end{equation}}
\newcommand{\bal}{\begin{align}}
\newcommand{\eal}{\end{align}}

\newcommand{\nn}{{\nonumber}}

\newcommand{\ket}[1]{\mbox{$ | #1 \rangle $}}
\newcommand{\bra}[1]{\mbox{$ \langle #1 | $}}

\newcommand{\beqa}{\begin{eqnarray}}
\newcommand{\eeqa}{\end{eqnarray}}

\newcommand{\tr}{\mathop{\mathrm{tr}}}
 
\definecolor{MyDarkBlue}{rgb}{0,0.08,0.45} 
\definecolor{MyLightMagenta}{cmyk}{0.1,0.8,0,0.1} 
\definecolor{MLM}{cmyk}{0.1,0.8,0,0.1} 
\definecolor{MyDarkGreen}{rgb}{0,0.45,0.08} 
\definecolor{MDG}{rgb}{0,0.55,0.05}

\usepackage[textsize=footnotesize]{todonotes}

\definecolor{atomictangerine}{rgb}{1.0, 0.6, 0.4}
\definecolor{bluegray}{rgb}{0.4, 0.6, 0.8}
\definecolor{brightube}{rgb}{0.82, 0.62, 0.91}
\definecolor{brilliantlavender}{rgb}{0.96, 0.73, 1.0}

\begin{document}

\title{
Light cone tensor network and time evolution
}

\begin{abstract}
The transverse folding algorithm [M. C. Bañuls \textit{et al.}, Phys. Rev. Lett. 102, 240603 (2009)] 
is a tensor network method to compute time-dependent local observables
in out-of-equilibrium quantum spin chains that can
overcome the limitations of matrix product states when entanglement grows slower in the time than in the space direction. 
We present a contraction strategy that makes use of the exact light 
cone structure of the tensor network representing the observables.
The strategy can be combined with the hybrid truncation proposed for global quenches in [Hastings and Mahajan, Phys. Rev. A 91, 032306 (2015)], 
which significantly improves the efficiency of the method.
We demonstrate the performance of this \emph{transverse light cone contraction} 
also for transport coefficients, and discuss how it can be extended to other dynamical quantities.
\end{abstract}

\author{Miguel Fr{\'{\i}}as-P{\'e}rez}\email{miguel.frias@mpq.mpg.de}
\affiliation{Max-Planck-Institut f\"ur Quantenoptik, Hans-Kopfermann-Str.\ 1, D-85748 Garching, Germany}
\affiliation{Munich Center for Quantum Science and Technology (MCQST), Schellingstr. 4, D-80799 M\"unchen}
\author{Mari Carmen Ba\~nuls}
\affiliation{Max-Planck-Institut f\"ur Quantenoptik, Hans-Kopfermann-Str.\ 1, D-85748 Garching, Germany}
\affiliation{Munich Center for Quantum Science and Technology (MCQST), Schellingstr. 4, D-80799 M\"unchen}
% Activate to display a given date or no date
%\date{\today}							
\maketitle

\section{Introduction}
\label{sec:intro}

Tensor networks (TN)~\cite{Verstraete2008,Schollwoeck2011,Orus2014a} have gained in the last decade a prominent role among numerical methods for  quantum many-body systems.
Simulating the dynamics of out of equilibrium systems remains nevertheless one of the most challenging open problems for these (and other) techniques.

In one dimensional systems, limitations of TN methods for dynamics are well understood: in global quenches the entanglement may grow fast~\cite{Calabrese2005,Osborne2006dyn,Schuch2008simul}, and the true state can escape the descriptive power of the TN ansatz. 
 This so-called entanglement barrier limits the applicability of the matrix product state (MPS)~\cite{Fannes1992,Vidal2003,Verstraete2004,PerezGarcia2007} description, and
 makes it difficult to predict the asymptotic long-time behavior, even when local observables in this limit are expected to be well-described by a thermodynamic ensemble,
 itself well approximated by a matrix product operator (MPO)~\cite{Verstraete2004a,Zwolak2004,Pirvu2010a,Hastings2006a,Molnar2015,Kuwahara2021thermal}.
A number of methods have been suggested to try to overcome this issue and extract information about the long-time behavior of local properties~\cite{Hartmann2009heisenberg,Banuls2009fold,Prosen2007,Muth2011,White2018therm,Rakovszky2020dissip,krumnow2019overcoming,Surace2019trading,Rams2020transport,Lopez2021broad,Kvorning2021}.
{While there is no universal solution, understanding the entanglement structures in the evolution TN can be crucial to identify the most adequate one for practical computations.} % in particular cases

\begin{figure}
	\centering
	\includegraphics[width=1.0\columnwidth]{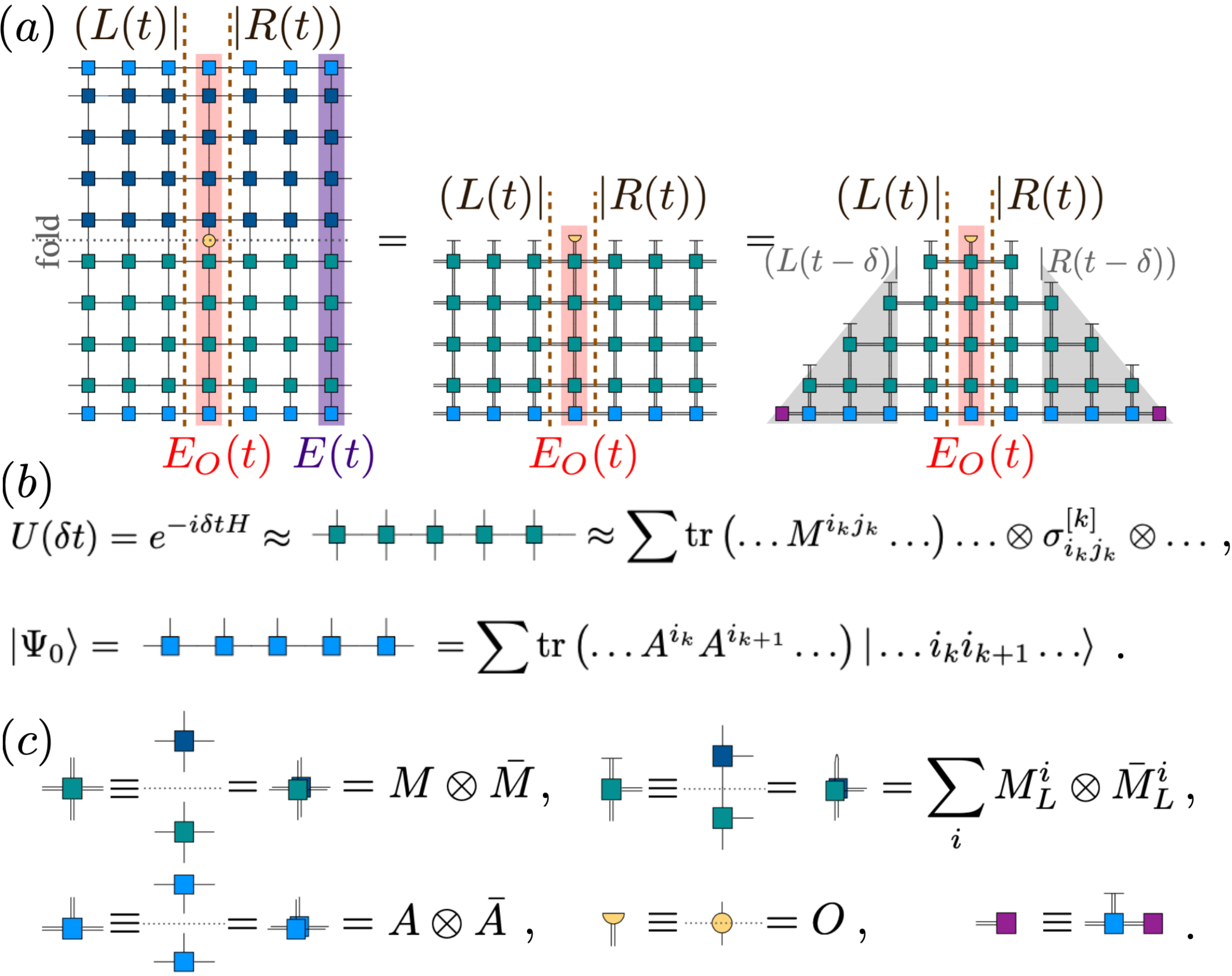}
	\caption{{(a, b)} Schematic construction of the minimal TN for the expectation value of a local operator $O$ after a global quench in a translationally invariant setting.
	At time $t=M \delta $ the expectation value $\bra{\Psi(t)}O\ket{\Psi(t)}$ corresponds to a two dimensional TN. After folding, the exact light cone is obtained after removing the mutually cancelling gates. 
	(c) Graphical notation for folded TN diagrams through the paper. }
	\label{fig:global_quench}
\end{figure}

In particular, the transverse folding strategy~\cite{Banuls2009fold,Mueller-Hermes2012,Hastings2015impro}
avoids the explicit representation of the evolved state as a MPS
and instead focuses on contracting a TN that represents exactly (up to Trotter errors)  the time-dependent observables.
Instead of the standard evolution in time direction, the folding algorithm contracts the TN along space.
In some scenarios, this allows local observables to be computed to longer times than other approaches~\cite{Banuls2011therm},
and it is an exact strategy for certain models~\cite{Piroli2020dual}.
Recently, there has been a rekindled interest in this approach,
triggered by the interpretation of the network in terms of an influence functional~\cite{Sonner2021inf,Lerose2021im,Ye2021influence}.

In local lattice models, the velocity of propagation of information is upper-bounded~\cite{Lieb1972,Hastings2006,Nachtergaele2006} and the exact TN for observables has a light cone structure.
While there have been proposals that exploit this fact to reduce the cost of the numerical simulation of the evolved state with TN~\cite{Hastings2009lc,Enss2012,Gillman2021finite,Zauner2015cmw,Phien2013dynwindow,Milsted2013nonuni},
and with quantum simulation~\cite{Haah2018quantum},
until now, the potential of combining it with the transverse strategy has not been
explored. 

Here we propose a strategy to exploit this property, a \emph{transverse light cone contraction} of the TN (\tlcc).
As in the original transverse folding, the \tlcc~ does not directly suffer from the entanglement growth in the state, and will be more efficient than standard algorithms when entanglement in the time direction grows slower than in the spatial one. But the \tlcc~ improves the efficiency with respect to the transverse folding in all cases, by reducing the computational effort to that of approximating the minimal network describing the time-dependent observables in a Trotterized evolution.
We demonstrate explicitly its performance for global quenches and different-time thermal correlators at infinite temperature, and investigate how the strategy can make use of the (more efficient) physical light cone determined by the Lieb-Robinson velocity~\cite{Hastings2006}.
We discuss possible extensions to other interesting quantities.

\begin{figure*}[ht]
	\centering
	\includegraphics[width=1.0\textwidth]{{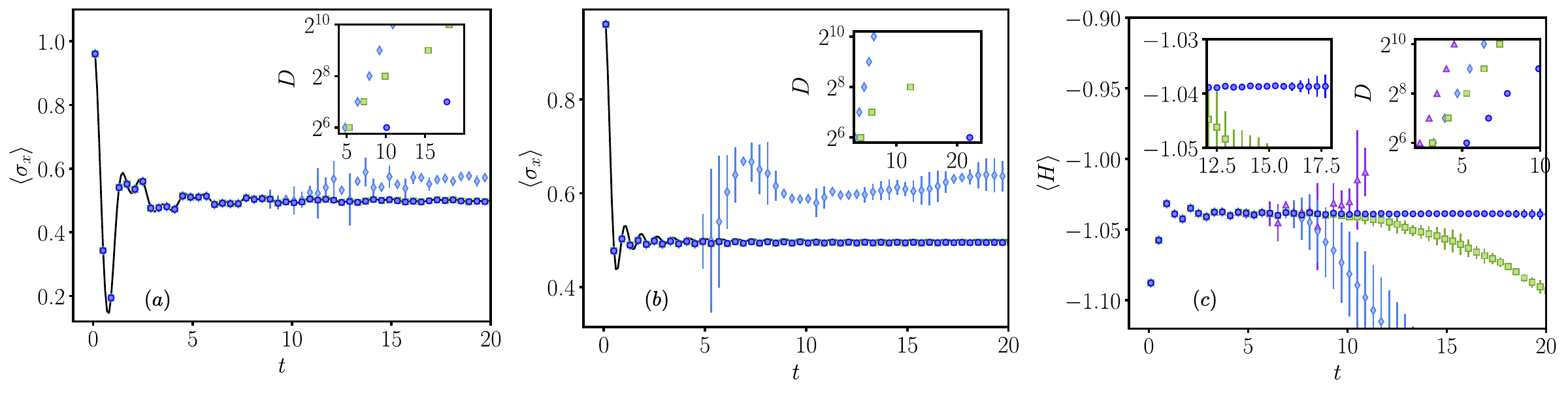}}
	\caption{
Evolution after a global quench from the initial state $\ket{X+}$, for the integrable [(a) $g=0.5$, (b) $g=1$] and non-integrable [(c) $g=-1.05$, $h=0.5$] Ising model. 
	The main plots show the transverse magnetization $\langle \sigma_x (t)\rangle$ (a, b)  and the energy density (c) computed with different algorithms using respectively  bond dimension $D=128$ (a,b) and $512$ (c). Error bars represent the difference with respect to the results obtained with $D'=D/2$. The TLCC contraction has been obtained both with the standard MPS truncation (green squares) and the
hybrid truncation of~\cite{Hastings2015impro} (dark blue circles).
For comparison, we also show the results of standard iTEBD (blue diamonds) and Heisenberg picture DMRG (purple triangles).\footnote{Heisenberg picture results are only shown in (c), since, in the integrable case, the operator in (a,b) can be exactly written as an MPO with constant bond dimension at all times \cite{Hartmann2009heisenberg}.}
 For the integrable case (a,b) we show also the analytic result (black line).
   The insets show the scaling of the bond dimension required to keep constant precision in each algorithm \cite{SM}. In the integrable case, this is compatible (at the later times) with a polynomial increase $D\sim t^{\alpha}$, consistent with observations in~\cite{Mueller-Hermes2012,Giudice2021te}. In the non-integrable case, the increase is compatible with an exponential growth for both truncation methods, but using the hybrid truncation exhibits a slower rate than standard ones, such that longer times can be reached with the same bond dimension. The left inset in (c) shows a zoom of the main plot to better appreciate the differences.
	}
	\label{fig:results_quench}
\end{figure*}

\section{Light cone tensor network for global quenches} 
\label{sec:quench}

The one-dimensional global quench is a natural test bench for time-evolution TN algorithms.
At time $t=0$ the system is prepared in a state that can be written as a MPS (e.g. a product state), and then it is let to evolve under a fixed Hamiltonian.
For simplicity, we restrict the discussion to a nearest-neighbour model, and a translationally invariant case, but the construction generalizes 
straightforwardly to any model with local (finite-range) interactions and some non-translationally invariant scenarios.

The transverse folding proposal of~\cite{Banuls2009fold}
starts from a two-dimensional TN whose contraction represents some time-dependent observable, such as a local expectation value. 
This TN can be constructed from a Suzuki-Trotter approximation of the evolution operator, where
the evolution for a discrete step of time $\delta$ can be approximated as a matrix product operator (MPO)~\cite{Verstraete2004a,Zwolak2004} with a small bond dimension, constructed from a product of two-body gates~\cite{Pirvu2010a}. 
The TN for the observable at time $t=M \delta$ is obtained by applying $M$ copies of this MPO with the initial state, which yields the evolved state, and contracting the operator of interest between this and its adjoint.

While standard TN algorithms as TEBD or tMPS~\cite{Vidal2004,Vidal2007infinite,Verstraete2004a,Paeckel2019tevolRev} compute the observable by contracting the network in the time 
direction, the transverse folding strategy performs the contraction in the spatial direction, after folding the TN in half,
such that tensors for the same site and time step in the ket and the bra are grouped together (see figure~\ref{fig:global_quench}a). 
After folding, the growth of entanglement in the time direction can be slower than in the spatial one, with the most dramatic difference observed for integrable systems~\cite{Mueller-Hermes2012,Giudice2021te}, but occurring also in generic cases, as the ones shown here.
When this difference in growth is present, the transverse strategy allows reaching longer times than standard algorithms.

For a translationally invariant system in the thermodynamic limit, the transverse contraction reduces to an expectation value of the form $(L(t)|E_O(t)|R(t))$, where $(L(t)|$ and $|R(t))$ are the
dominant left and right eigenvectors of the transfer operator $E(t)=\sum_i \bar{A(t)}^i \otimes A^i(t)$, and $E_O(t)=\sum_i \bar{A(t)}^i \otimes A^j(t) \langle i|O|j\rangle$~\cite{PerezGarcia2007}. Here, $A^i(t)$ represents the \emph{concatenated}~\cite{Hubener2010concatenated} local tensor of the time-dependent state, itself a MPO.
In the transverse folding strategy, the boundary vectors $(L(t)|$ and $|R(t))$ are approximated by MPS.
This approximation can be found, for instance, via a power iteration or a Lanczos algorithm, using repeated MPO-MPS contractions.

Such strategies do not take into account that the TN has a light cone structure.
Because the individual gates are local, outside the causal cone of the
operator, each gate cancels with its adjoint. This ensures that each of the required boundary vectors (dominant eigenvectors of the 
transfer operator) corresponds precisely to the contraction of a triangular network as depicted in fig.~\ref{fig:global_quench}.
We can approximate directly the contraction of such triangle in the space direction by a MPS.
 This strategy, which we call \emph{transverse light cone contraction} (\tlcc), allows us to obtain $(L(t)|$ and $|R(t))$ in a fixed number of steps (proportional to $M$).
Furthermore, once we have found the vectors for $M$ time steps, we can directly 
obtain them for $M+1$ by applying a single MPO (as illustrated in the figure), which increases the length by one, and 
approximating the result via a single truncation step.
This step can be performed using standard MPS truncation algorithms, which reduce the bond dimension by minimizing 
a distance between the truncated vector and the original one. However, for this particular problem the \emph{hybrid} truncation algorithm proposed in~\cite{Hastings2015impro}, which effectively \emph{evolves} the bond of the boundary vector according to the real time dynamics, yields a much more efficient use of the available bond dimension (see also insets of fig.~\ref{fig:results_quench}).

The \tlcc~strategy results in a more efficient algorithm than the originally proposed folding, which required iterative MPO-MPS 
contractions until convergence of the dominant eigenvectors, run independently for each different time step (in particular, for the cases analyzed in this work, we find the power iteration required several tens of MPO-MPS contractions per time step).
Notice, nevertheless, that if the bond dimension used is large enough, both the original folding algorithm and the \tlcc~should result in the same boundary vector.
What ultimately determines the applicability of transverse strategies is thus the amount of \emph{entanglement}
present in the transverse network. 

To probe the performance of the method, we consider a quantum Ising chain, initialized in a 
product state $\ket{X+}=\lim_{N\to\infty}[(\ket{0}+\ket{1})/\sqrt{2}]^{\otimes N}$.
We then apply the Hamiltonian,
\beq
\His=\sum_{i}\left(J\sigma_i^{z} \sigma_{i+1}^{z}+g \sigma_i^{x}+h\sigma_i^{z}\right ),
\label{eq:Hising}
\eeq
and compute local expectation values after time evolution. 
In all the following we fix $J=1$, and a Trotter step $\delta=0.1$, and 
vary the parameters of the model to study integrable ($g=\{0.5,\,1\}$, $h=0$) and non-integrable ($g=-1.05$, $h=0.5$) regimes. 
Figure~\ref{fig:results_quench} shows the results and demonstrates that the \tlcc~can efficiently simulate the integrable
quenches. In the non-integrable regime, the required bond dimension grows much faster with time, but
the method is still advantageous as compared to standard evolution, much more so when the truncation is performed as in~\cite{Hastings2015impro} (see right inset of fig.~\ref{fig:results_quench}c).

\section{Light cone tensor network for transport coefficients} 
\label{sec:transport}

\begin{figure*}[ht]
	\centering
	\includegraphics[width=1.0\textwidth]{{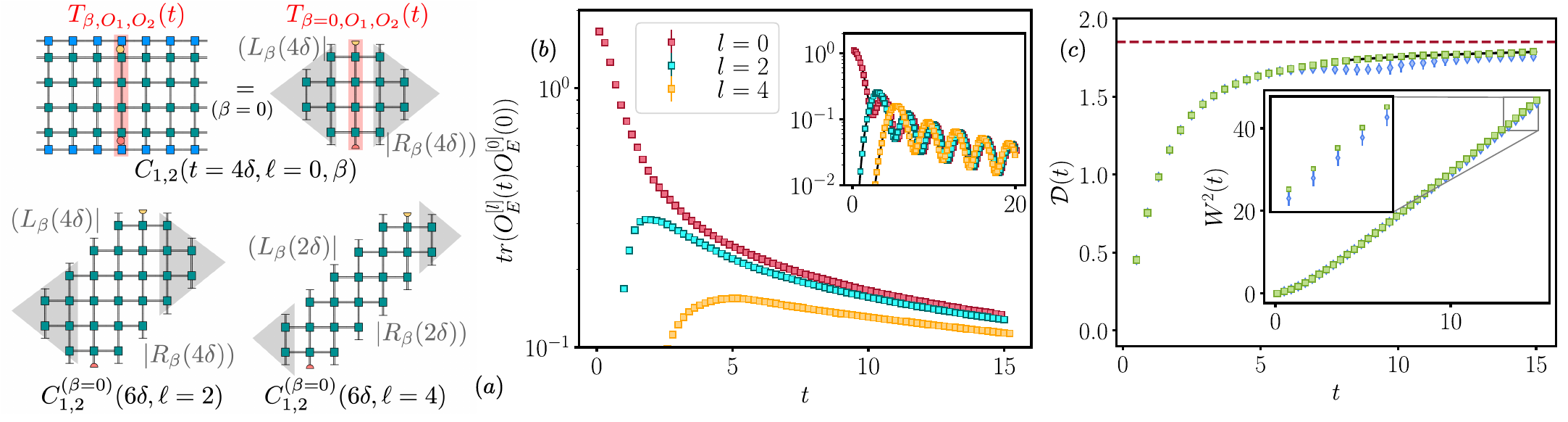}}
	\caption{(a) Schematic construction of the minimal TN for two-point correlators at infinite temperature for different times and distances \cite{SM}.
	(b) Energy autocorrelations $C_{EE}(t,\ell,\beta=0)$ obtained from the TLCC method at several distances as a function of time in the integrable ($g=0.5$, inset) and non-integrable ($g=-1.05$, $h=0.5$, main plot) Ising chain at $\beta=0$. The error bars (smaller than the size of the marker) show the difference between results with two different bond dimensions ($D$,$D'$) [for the inset (200, 100), for main plot (500, 200)]. In the inset, the black curves
	represent the results coming from the analytical solution of the model.
(c) Spatial variance~\ref{eq:variance} of the normalized autocorrelations~\eqref{eq:variance} (inset) and corresponding diffusion constant (main plot) in the non-integrable case obtained from the TLCC (green squares)  and TEBD (blue diamonds) with $D=1024$, with error bars showing the difference with respect to $D'=512$. The solid black line in the main plot shows a fit of the form $\mathcal{D}_E \exp(b/t)$, which predicts the asymptotic value $\mathcal{D}_E\approx1.9$ (red dotted line). 
	}
	\label{fig:correlator}
\end{figure*}

The same idea can be adapted to the computation of other dynamical quantities.
It is the case of thermal correlators, of the  form $C_{1,2}(t,\ell,\beta)=\tr (\rho_{\beta} O_2^{[\ell]}(t) O_1^{[0]}(0))$,
where $\rho_{\beta}=e^{-\beta H}/Z$ is the thermal equilibrium state at inverse temperature $\beta$, $Z=\tr(e^{-\beta H})$ is the partition function, $O_{k}^{[\ell]}(t)$ is a (local) operator acting on site $\ell$ at time $t$, 
and $O_k(t)=U(t)^{\dagger} O_k U(t)$ is the time-evolved operator in Heisenberg picture. 
Since $[\rho_{\beta},H]=0$, the thermal state is invariant under the evolution, and using
$\rho_{\beta}\propto \rho_{\beta/2} \rho_{\beta/2}^{\dagger}$  
we can write (up to normalization),
$C_{1,2}(t,\ell,\beta) \propto \tr (U(t)^{\dagger} \rho_{\beta/2}^{\dagger} O_2^{[\ell]} U(t) O_1^{[0]} \rho_{\beta/2})$. 
Using a MPO approximation to $\rho_{\beta/2}$ (obtained with standard TN methods~\cite{Verstraete2004a,Zwolak2004,Feiguin2005temp,Chen2018xtrg}), and the Trotterized real time evolution as in the previous section, this quantity can be expressed as a two 
dimensional folded TN, which can be contracted in the temporal~\cite{Barthel2012thermal,Karrasch2012finiteT,Barthel2013thermal}
or spatial (transverse)~\cite{Mueller-Hermes2012} direction.

Due to the invariance of the thermal state, each local observable 
generates also a light cone structure that can be exploited in the \tlcc~approach.
Now the cancellation of gates outside the causal cone of the operators occurs both at the upper and the lower parts of the network 
(see figure~\ref{fig:correlator}a), and the minimal TN has a rectangular form, resembling a pillow,
a structure which was used in~\cite{Suenderhauf2018} to evaluate correlators in random quantum circuits.
The \tlcc~strategy again requires contracting a triangular TN corresponding to the 
lateral corners of the figure to obtain boundary vectors $(L_{\beta}(t)|$ and $|R_{\beta}(t))$.
\footnote{Different from the global quench above, in this case each iteration of the algorithm grows the boundary vectors in two time steps.}
If both operators act on the same site ($\ell=0$), the time dependent correlators can be expressed as a contraction $(L_{\beta}(t)|T_{\beta,O_1,O_2}(t) |R_{\beta}(t))$, with a single MPO $T_{\beta,O_1,O_2}(t)$ constructed from concatenating the local tensors for the unitaries, the operators and the states (see fig.~\ref{fig:correlator}a).  
For correlators at non-zero distance $\ell$ the minimal TN becomes elongated (fig.~\ref{fig:correlator}a, lower diagrams). To approximate its contraction, the boundary vectors $(L_{\beta}(t)|$ and $|R_{\beta}(t))$ for a certain time $t$ are first grown to incorporate, respectively, $O_1$ at the bottom of the TN, and $O_2$ at the top. These extended vectors contain the evolution steps up to time $t+2 \delta $, and can be contracted together to obtain the correlators at $\ell=1$ for times $t+3\delta $ and $t+4\delta $.
The vectors can be then evolved again, following the TN structure, 
which does not increase their length, but allows access to correlators at any later time $t+(2+k)\delta $ and distances $\ell=k,\, k+1$.
Applying this systematically we can obtain all non-vanishing correlators. 
This generalizes trivially to operators on more than one site, or with MPO structure.

Here we illustrate the simplest case, infinite temperature, where $\rho_{\beta=0}\propto \Id$ 
and the contour of the TN becomes uncorrelated. 
We consider the energy density operator 
\beq
O_E^{[i]}:=J\sigma_i^{z}\sigma_{i+1}^{z}+\frac{g}{2}(\sigma_i^{x}+\sigma_{i+1}^{x})+\frac{h}{2}(\sigma_i^{z}+\sigma_{i+1}^{z}),
\eeq
which can be written as a MPO of range 2.
Figure~\ref{fig:correlator}b shows our results for the correlators $C_{EE}(t,\ell,\beta=0)$ as a function of time for several distances in the non-integrable  ($g=-1.05$, $h=0.5$, main plot) and integrable ($g=0.5,\, h=0$, inset) cases \cite{SM}. 

Specially interesting is the possibility of ab initio calculations of transport properties~\cite{Bertini2021review_transport} in non-integrable models.
In particular, diffusion constants can be related to the spatial spreading in time of autocorrelations of a density~\cite{Steinigeweg2009spread,Kim2013ballistic,Rakovszky2020dissip}.
Normalizing the correlators as $\tilde{C}_{EE}(0,\ell):=C_{EE}(t,\ell)/\sum_{\ell}C_{EE}(0,\ell)$,
a diffusion constant $\mathcal{D}(t)$ may be obtained from their spatial variance~\cite{Steinigeweg2009spread} 
 \beq
 W^2(t):=\sum_{\ell} \tilde{C}_{EE}(t,\ell) \ell^2-\left( \sum_{\ell} \tilde{C}_{EE}(t,\ell) \ell \right)^2,
 \label{eq:variance}
 \eeq
 as $\frac{\partial W^2}{\partial t}=2 \mathcal{D}(t)$.
Figure~\ref{fig:correlator}c shows the (linearly growing) variance $W^2(t)$ (main plot), and the corresponding diffusion constant (inset) obtained from the correlators
for the non-integrable case.
The diffusion constant is well fitted by a function $\mathcal{D}(t)=\mathcal{D}_E \exp(b/t)$, compatible with saturation to a constant $\mathcal{D}_E\approx 1.9$ in the asymptotic regime \footnote{The fit $D(t) = \mathcal{D}_E \exp(b/t)$ is a heuristic choice that describes our data well over a range of fitting windows and allows us to extrapolate to the limit of infinite time. We have also tried successfully fits with polynomials of 1/t , and found compatible results.}.
While TEBD (blue diamonds) produces close values for the same quantities, the error is appreciable in the diffusion constant already at short times.

\section{The physical light cone}
\label{sec:LR}

\begin{figure}
	\centering
	\includegraphics[width=1.0\columnwidth]{{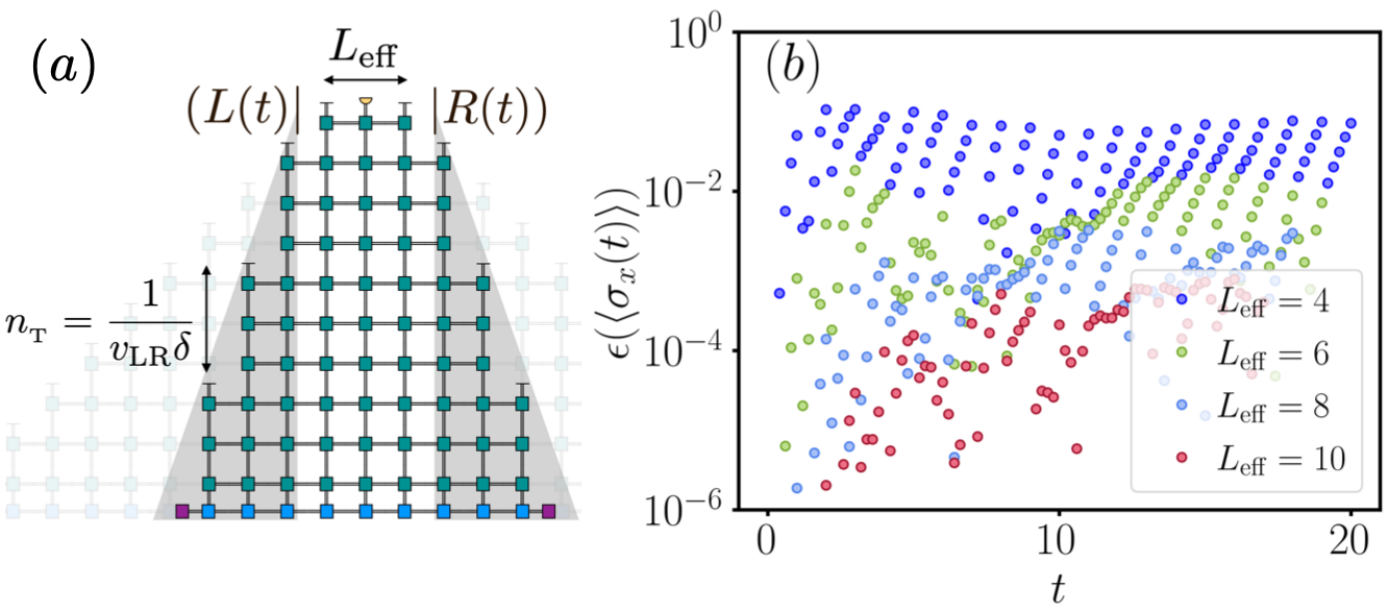}}
	\caption{(a) The physical velocity defines a much narrower light cone than the Trotterization (background). (b) Relative difference between $\langle \sigma_x \rangle$ computed with the LR and Trotter light cones for the integrable global quench of fig.~\ref{fig:results_quench}a, for which $n_{\mathrm{T}}=10$ and different sizes of the subsystem $L_{\mathrm{eff}}$, with $D=200$ in all cases. }
	\label{fig:resultsLR}
\end{figure}

In general, we expect that the physical light cone is much narrower than the trivial one from the Trotterization,  used in the previous sections.
We could thus approximate the TN by a light cone one in which the slope corresponds to the maximal physical velocity $v_{\mathrm{LR}}$.
This can be achieved by implementing a more efficient \tlcc~growing iteration, in which $n_{\mathrm{T}}=1/(v_{\mathrm{LR}} \delta)$ time steps are applied at once every time a space site 
is contracted (fig.~\ref{fig:resultsLR}a). Notice that this light cone is not exact, but has (exponential) corrections. 
Thus it is convenient to 
consider the light cone for a subsystem of size $L_{\mathrm{eff}}$  that includes the support of the operator.\footnote{Equivalently, we can \emph{insert} in the middle of the column, 
corresponding to an earlier time.} 

To probe this reduced light cone we choose an integrable instance,
($g=0.5,$ $h=0$), for which the Lieb-Robinson velocity is known ($v_{\mathrm{LR}}=1$, corresponding to $n_{\mathrm{T}}=10$ with our Trotter step),
and simulate the global quench of fig.~\ref{fig:results_quench}a.
Compared to \tlcc~ for the full light cone with the same bond dimension,
we observe (fig.~\ref{fig:resultsLR}) that the physical one, determined by $v_{\mathrm{LR}}$, captures indeed 
the correct evolution: while the narrower light cone deviates from full results, the errors are reduced exponentially 
(until the level of original truncation error) by considering a small window $L_{\mathrm{eff}}$.

\section{Discussion}
\label{sec:discussion}

We have presented a strategy that builds on the transverse folding~\cite{Banuls2009fold} to approximate time-dependent observables in a one-dimensional quantum system.
Noticing the exact light cone structure of the TN and implementing its transverse contraction, it is possible to compute long time properties in a more efficient manner.
Combined with the hybrid truncation~\cite{Hastings2015impro}, 
this allows us to reach longer times with a smaller bond dimension whenever the temporal entanglement grows slower than the physical one, which, as we have seen, happens not only for integrable systems. 
It is possible to use the physical upper bound of the Lieb-Robinson velocity to further restrict the width of the relevant TN and define a more efficient iteration.

We have evaluated the performance of the \tlcc~strategy for integrable and non-integrable global quenches, and for transport 
properties at infinite temperature.
With minimal changes, the method extends to other scenarios, such as finite temperature or non translationally invariant setups including impurities or a contact between two chains.
It is furthermore possible to adapt the strategy to other more complex dynamical quantities.

The basic \tlcc~does not require additional hypothesis to truncate observables or states. 
Its convergence can be systematically explored as the bond dimension is increased. 
What ultimately limits the validity of the strategy is the entanglement in the 
time direction, which strongly depends on the setup and the model~\cite{Mueller-Hermes2012,Lerose2021integ,Giudice2021te}. 
The behavior of the \tlcc~can thus provide useful information to determine
optimal strategies for different problems.
Another parameter in the approximation is the Trotter step, which is known to affect the entanglement growth in standard algorithms~\cite{Paeckel2019tevolRev}. Since simulations with different $\delta$ may be necessary to extrapolate the exact results, it is also interesting to study how varying $\delta$ affects our observations.
Further interesting avenues for future investigation are exploring the TN cut according to different velocities, 
to explore the propagation of correlations in the TN and effectively measure $v_{\mathrm{LR}}$.

While we were completing this manuscript, an equivalent strategy for global quenches was independently suggested in~\cite{lerose2022overcoming}.

\acknowledgments

We are thankful to J. I. Cirac, M. Hastings and L. Tagliacozzo for insightful discussions at different stages of this project.
This work was partly supported by the Deutsche Forschungsgemeinschaft (DFG, German Research Foundation) under Germany's Excellence Strategy -- EXC-2111 -- 390814868.
M.C.B. acknowledges the hospitality of KITP, where earlier versions of the work were developed, with support from the National Science Foundation under Grant No. NSF PHY-1748958.

\bibliographystyle{apsrev4-1}
\bibliography{lightconeTN}

\pagebreak
\onecolumngrid
\vspace{\columnsep}
\newpage
\begin{center}
\textbf{\large Supplementary Material: Light cone tensor network and time evolution}
\end{center}
%\vspace{2cm}
\twocolumngrid

\section{Tensor networks for thermal response functions}

\begin{figure}[ht]
	\centering
	\includegraphics[width=.9\columnwidth]{{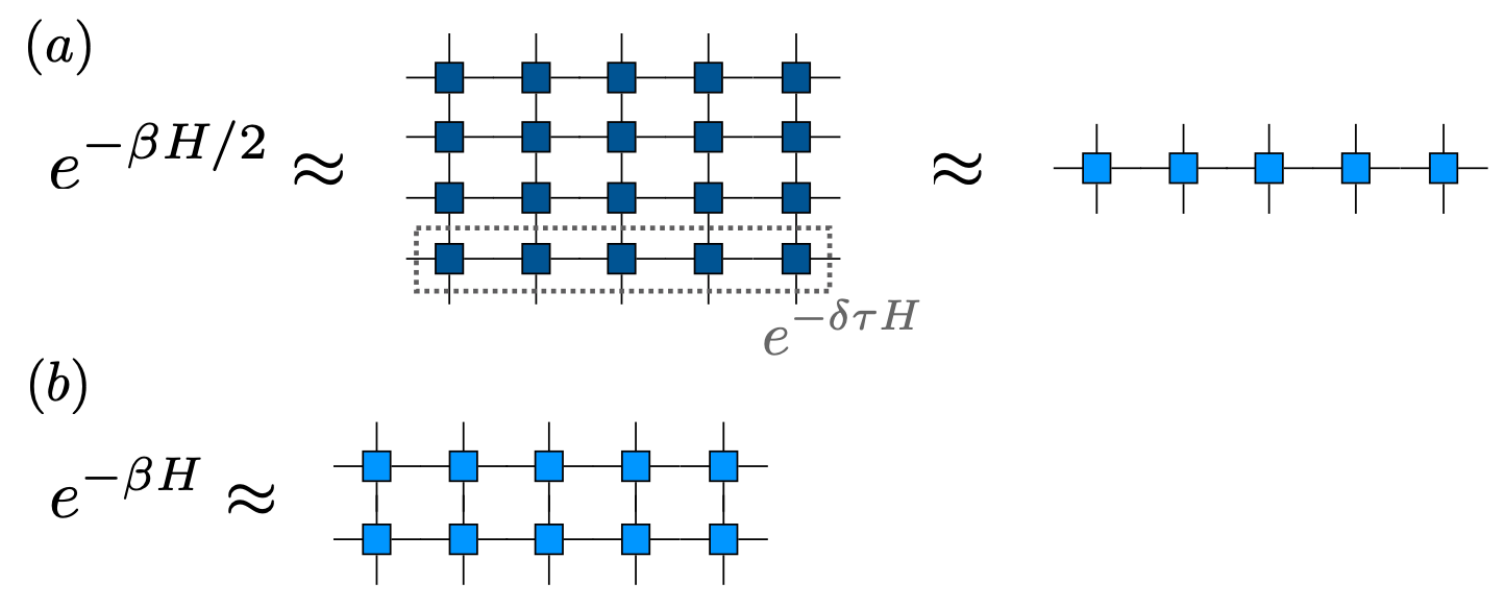}}
	\caption{ Schematic description of a TN algorithm to find the MPO approximation of the Gibbs ensemble. (a) The MPS approximation of the thermofield state found by iteratively applying small imaginary time steps onto the maximally entangled state is equivalent to an MPO approximation of the exponential operator. (b) Tracing out the ancillary degrees of freedom is equivalent to considering $(e^{-\beta H/2})^{\dagger} e^{-\beta H/2}$, which is guaranteed to be positive.}
	\label{figS:rhoBeta}
\end{figure}

\begin{figure}[ht]
	\centering
	\includegraphics[width=.98\columnwidth]{{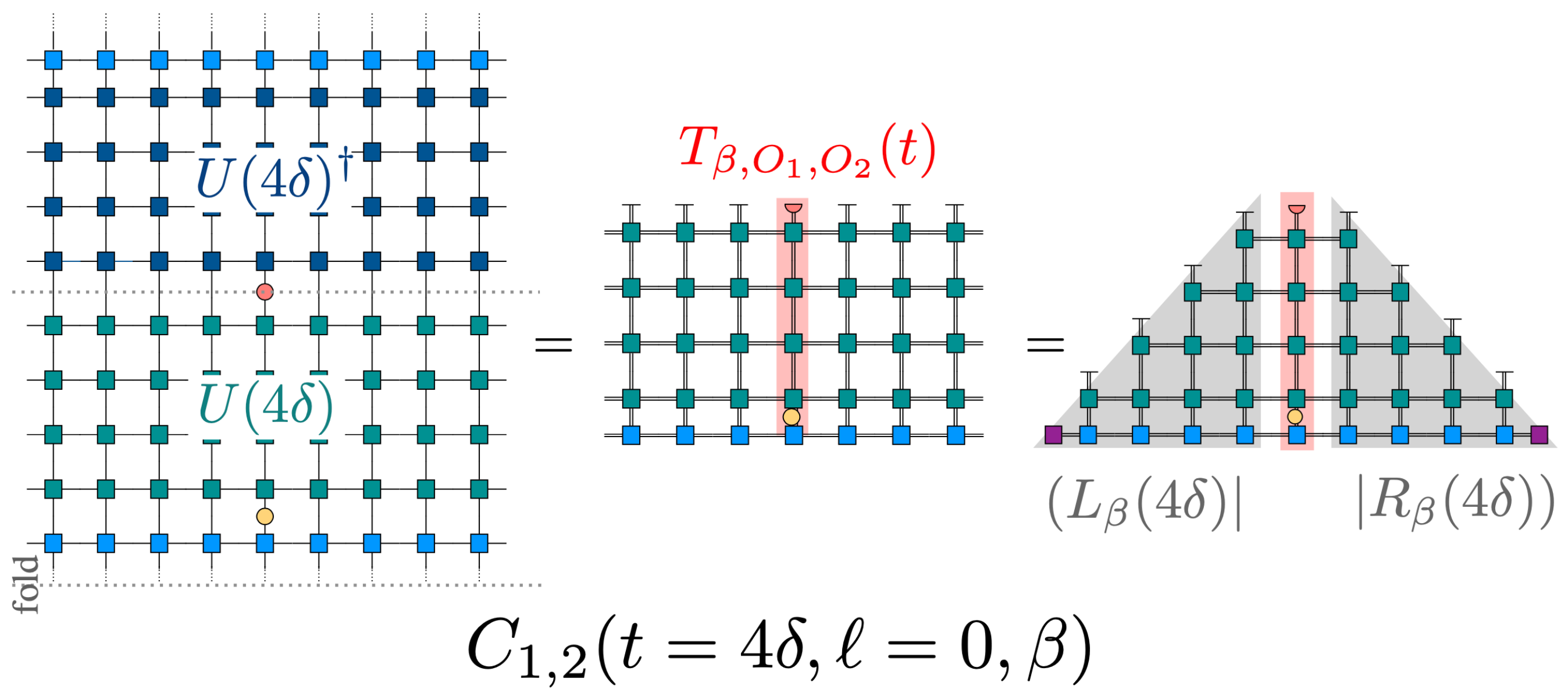}}
	\caption{TN for the thermal correlators. The left diagram shows the network for the observable~\eqref{eqS:C12} (the open dotted lines above are connected to the corresponding ones on the lower edge). A folded version across the dotted horizontal line, shown in the middle, can be written in terms of the same double tensors defined in Fig. 1 of the main text with the thermal MPO tensors defining the boundary. For local $O_2$, local unitaries cancel out, resulting in the light cone structure on the right, analogous to the one for pure state quenches.}
	\label{figS:C12}
\end{figure}

\begin{figure}[ht]
	\centering
	\includegraphics[width=.95\columnwidth]{{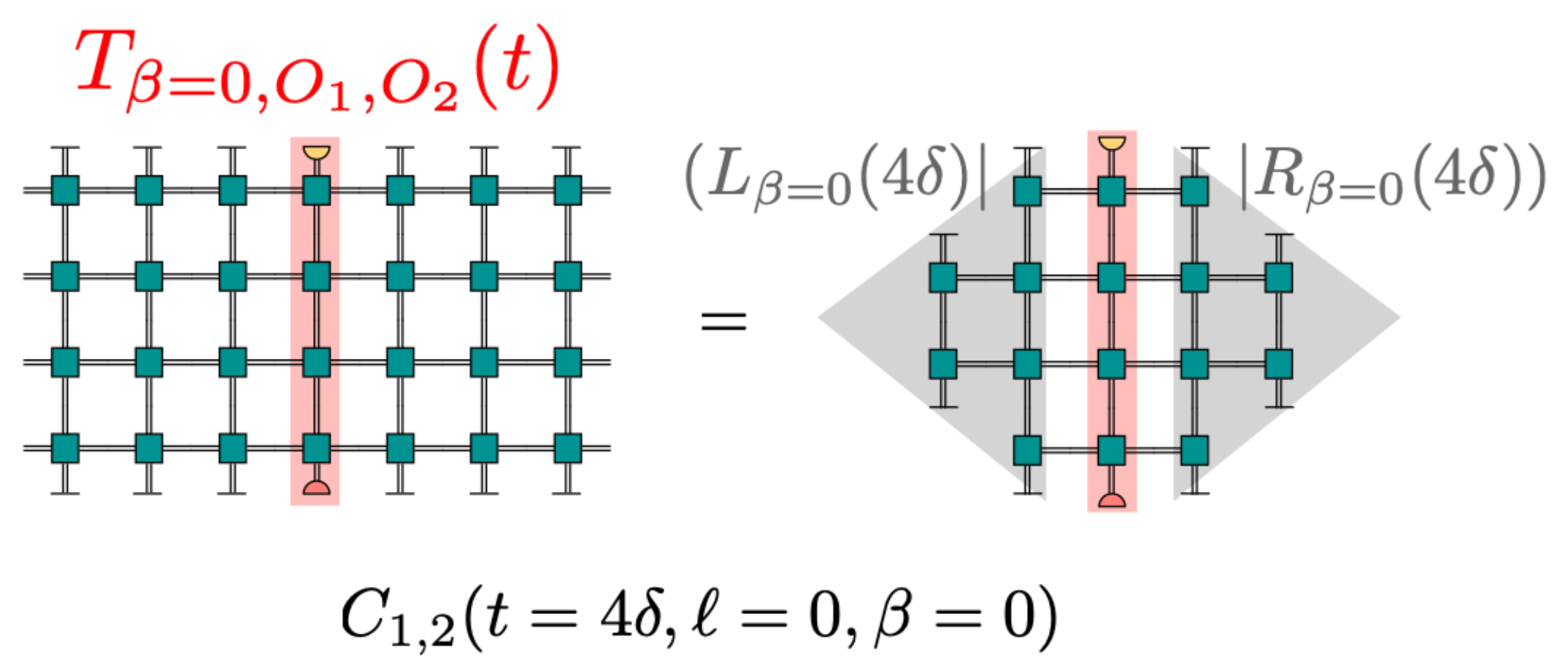}}
	\caption{TN for the thermal correlators at infinite temperature. In the particular case $\beta=0$, the thermal MPO is exact and proportional to identity, and local unitaries cancel around local $O_1$, resulting in a diamond shape.}
	\label{figS:beta0}
\end{figure}

\begin{figure}[ht]
	\centering
	\includegraphics[width=.95\columnwidth]{{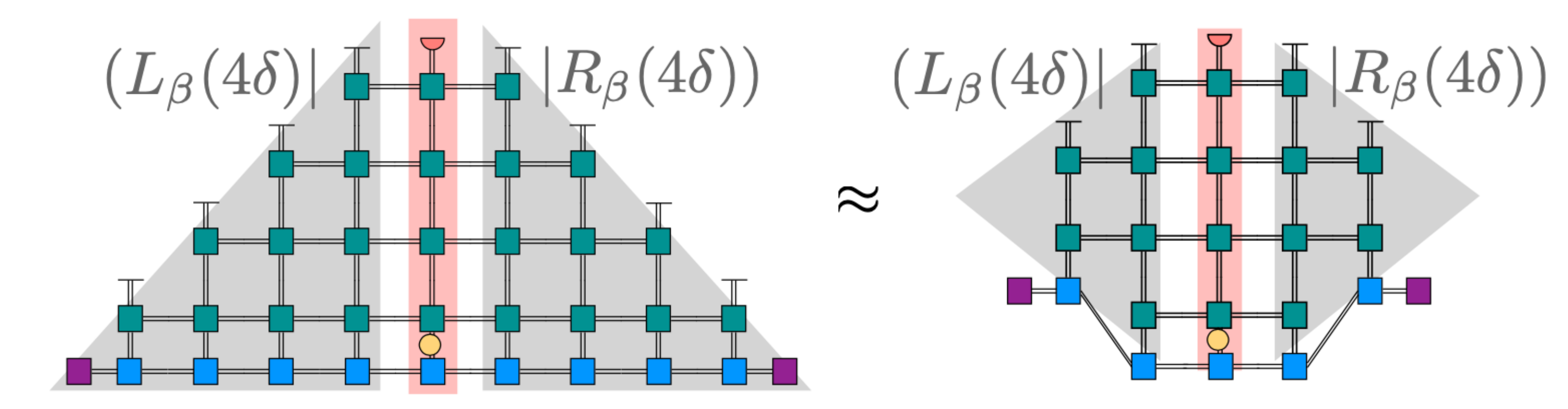}}
	\caption{TN for the thermal correlators at finite temperature. In the general case, the thermal MPO is not exact, and the diamond-shaped TN will approximate the full one.}
	\label{figS:beta}
\end{figure}

\begin{figure}[ht]
	\centering
	\includegraphics[width=.98\columnwidth]{{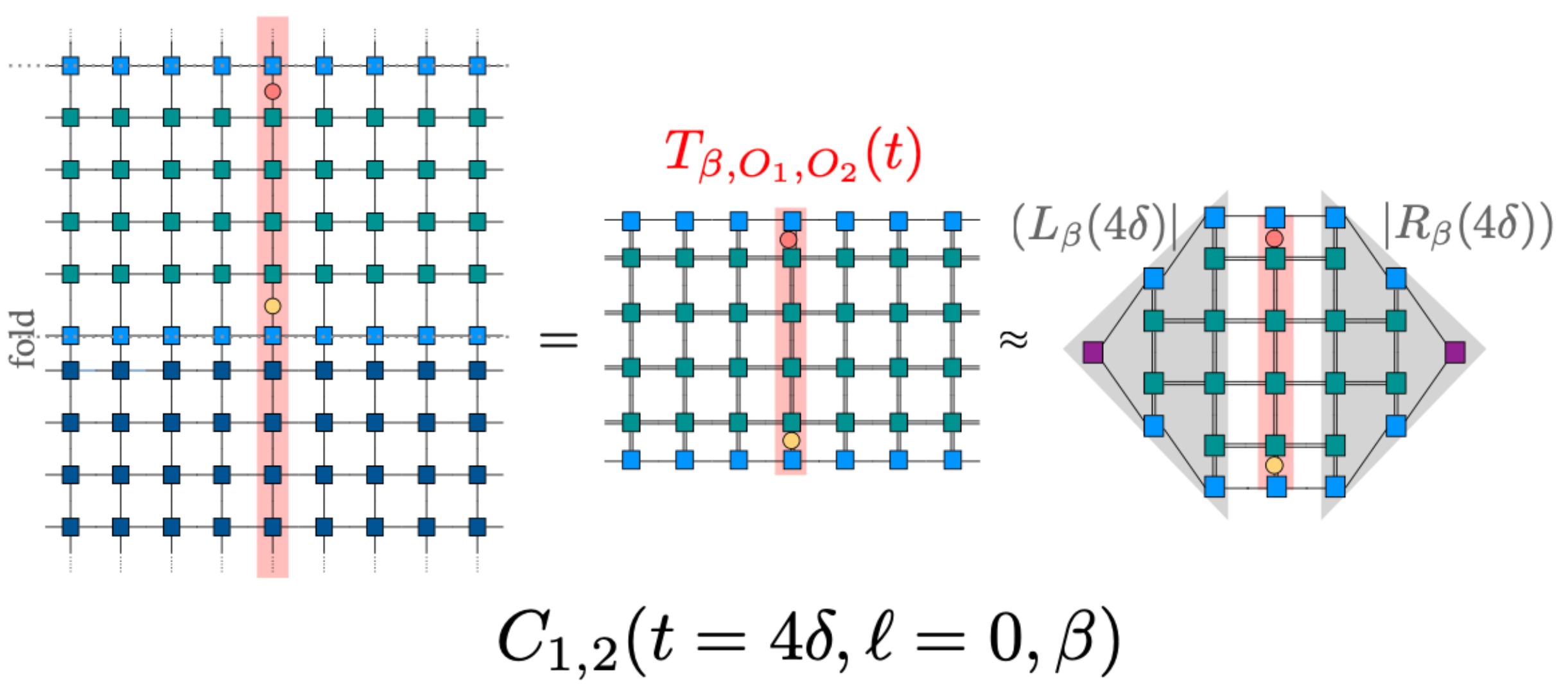}}
	\caption{Alternative construction of the TN for the thermal correlators using the purification structure. }
	\label{figS:C12_puri}
\end{figure}

This section shows explicitly how to construct the TN of Fig.~3(a) in the main text, for the case of arbitrary inverse temperature $\beta$.
We are interested in correlators of the form
$C_{1,2}(t,\ell,\beta)=\tr (\rho_{\beta} U(t)^{\dagger} O_2^{[\ell]} U(t) O_1^{[0]})$,
where $\rho_{\beta}=e^{-\beta H}/Z$ is the thermal equilibrium state at inverse temperature $\beta$, $Z=\tr(e^{-\beta H})$ is the partition function, $O_{k}^{[\ell]}$ is a (local) operator acting on site $\ell$, 
and $U(t)=e^{-i H t}$ is the time-evolution operator for time $t$.
In order to write this quantity as the contraction of the TN shown in the text, we start by finding an MPO approximation of the thermal state. This can be achieved with several algorithms~\cite{Verstraete2004a,Zwolak2004,Feiguin2005temp,Chen2018xtrg}. Here we illustrate (see Fig.~\ref{figS:rhoBeta}) the (possibly) most common algorithm, based on a purification and a Trotter expansion of $\rho_{\beta}$.

The purification approach is equivalent to considering a thermofield double state, i.e. a pure state of the form
\beq
\ket{\Psi_{\beta}}\propto e^{-\beta H/2} \ket{\Phi},
\eeq
where $\ket{\Phi}$ is a maximally entangled state of the system and an ancillary copy of it.
Tracing out the ancillary system results (up to normalization) in the Gibbs ensemble $\rho_{\beta}$.
For any basis $\{\ket{n}\}$ of the system, we can write
\beq
\ket{\Psi_{\beta}}\propto e^{-\beta H/2} \sum_n \ket{n,n}.
\label{eqS:thermof}
\eeq
The most frequently used TNS algorithm for thermal equilibrium states proceeds by approximating $\ket{\Psi_{\beta}}$ by an MPS in a basis in which each system site is grouped with an ancillary one (forming effective sites of dimension $d^2$). 
To do so, the state is initialized to the maximally mixed one between system and ancilla (equivalent to the vectorized identity operator), i.e. a MPS with bond dimension one. 
The exponential operator $e^{-\beta H/2}$ can be discretized as the product of a finite number $M=\beta/(2 \delta \tau)$ of imaginary time steps of length $\delta \tau$.
If the Hamiltonian is local, each of them, can be approximated by a MPO (for instance, for nearest-neighbor models, one can use an even-odd Trotter approximation) and successively applied on the MPS, acting on the system degrees of freedom. After each step, a standard truncation can be performed (using any of the algorithms for Trotterized time evolution), such that after a fixed number of steps $M=\frac{\beta}{2 \delta \tau}$, an MPS approximation to~\eqref{eqS:thermof} is obtained.
As sketched in Fig.~\ref{figS:rhoBeta}(a), this procedure is equivalent to approximating the exponential by an MPO (with the Frobenius norm  characterizing the quality of the approximation in the operator level).
Whereas this procedure could be run for the full inverse temperature, to obtain an MPO approximation of $e^{-\beta H}$, the truncation in MPO-MPS products does not preserve positivity. Instead, an MPS approximation of the thermofield state results necessarily in a positive density matrix (with purification structure) when tracing the ancillas,
as shown schematically in Fig.~\ref{figS:rhoBeta}(b).

The TN for the thermal response functions can be constructed applying the operator $O_1$ followed by the Trotterized time evolution on this MPO, and finally applying $O_2$ (possibly on a different site) and taking the trace. This results in a TN that is periodic in the time direction. Folding (or flattening) it results in a doubled TN, similar to the one obtained for pure state evolution, as illustrated in Fig.~\ref{figS:C12}.

The simplest case is that of infinite temperature ($\beta=0$), when the thermal state has an exact MPO representation with bond dimension one, since $\rho_{\beta=0}\propto \Id$. Then the local unitary matrices that represent the real time evolution cancel also around $O_1$ exactly, and the TN to be contracted has a diamond shape (see fig.~\ref{figS:beta0}).

At arbitrary temperature, the cancellation around $O_1$ is no longer exact, since the MPO representation of the thermal state is only approximate, and local unitaries do not commute exactly with it. Thus we can consider the diamond-shaped TN in this case to be an approximation of the infinite one (see Fig.~\ref{figS:C12_puri}).  
We expect this to introduce a small error.
Notice that in more standard evolution algorithms (i.e. those which evolve the MPS in real time), exploiting this light cone structure has also been shown to be useful in infinite systems, 
for instance by considering an expanding window embedded in an infinite MPS~\cite{Zauner2015cmw,Phien2013dynwindow,Milsted2013nonuni}.

Finally, notice that the TN construction described above is not unique. 
If we make use of the property
\begin{align}
U O_1 e^{-\beta H} U^{\dagger}&=U O_1 e^{-\beta H/2} U^{\dagger} U e^{-\beta H/2} U^{\dagger}
\nn \\
&=U O_1 e^{-\beta H/2} U^{\dagger} e^{-\beta H/2},
\label{eqS:puriEvol}
\end{align}
which has been previously exploited for the simulation of real time evolution of thermal states~\cite{Barthel2012thermal,Karrasch2012finiteT,Barthel2013thermal}, and apply also the cyclic property of the trace, we can write the quantity of interest as (up to a normalization factor) \beq
C_{1,2}(t,\ell,\beta)\propto \tr \left(e^{-\beta H/2} O_2 U(t) O_1 e^{-\beta H/2} U(t)^{\dagger} \right).
\label{eqS:C12}
\eeq
This results in a different TN, illustrated in  Fig.~\ref{figS:C12_puri}, in which the upper and lower boundaries are given by the purification tensors. For infinite temperature, both constructions are equivalent, but for the general case, they will give rise to different approximation errors. A systematic analysis of these alternatives, the approximation error and its dependence on temperature will be carried out elsewhere.

\section{Error estimates for the different approaches}

In the insets of Figure 2 of the main text, we show the scaling of the bond dimension needed with time to maintain fixed precision in the global quench scenario. Here in this appendix, we provide some extra information on how the scaling is computed and what we mean by fixed precision. 

\begin{figure}[]
	\centering
	\includegraphics[width=.98\columnwidth]{{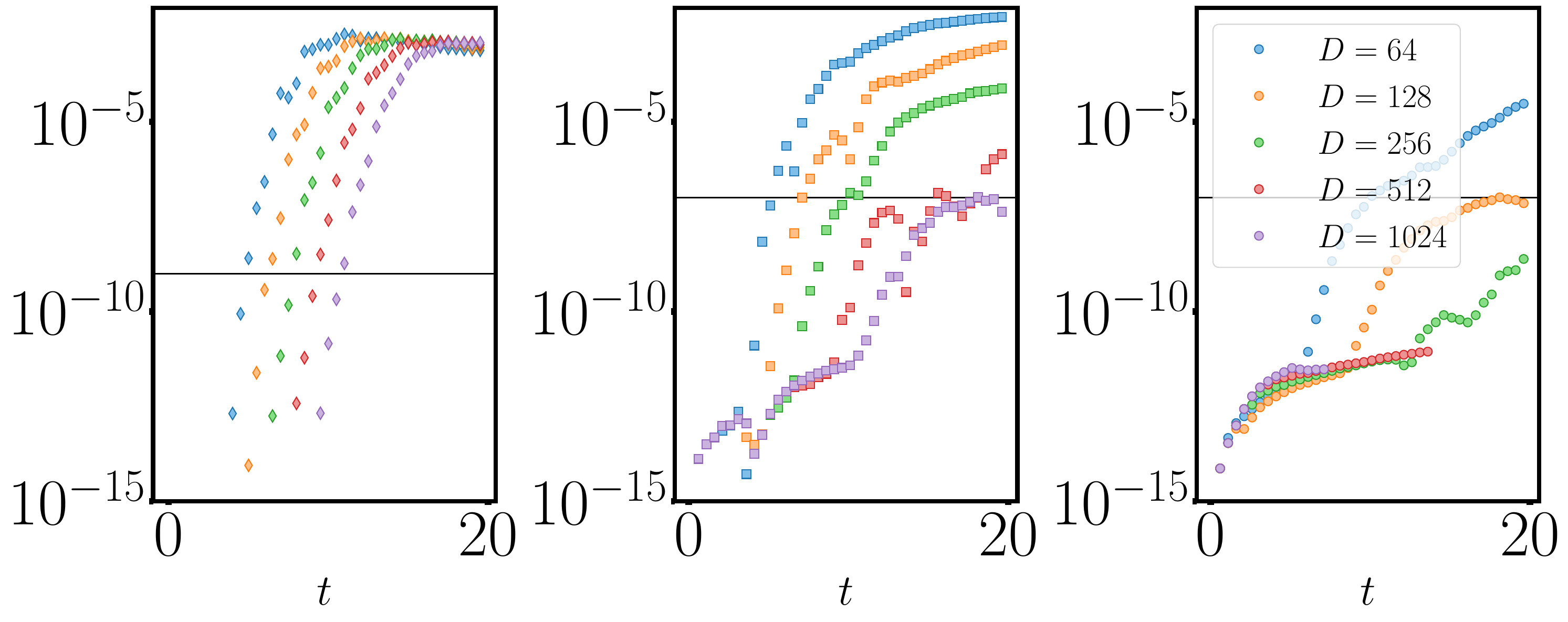}}
	\caption{Scaling of the error measures for the different algorithms in the non-critical integrable case $(J = 1, g = 0.5, h = 0)$. The left plot shows the growth of the sum of the squares of the discarded Schmidt weights for the different bond dimensions simulated with TEBD, and the center and right plot show the deviation of the expectation value of the identity from one as a function of time for the TLCC algorithm with the standard MPS truncation (center) and hybrid truncation from \cite{Hastings2015impro} (right). The black lines in each plot show the constant level of error we use to make the scaling in the insets of Fig.2 of the main text.}
	\label{figS:errorScalingInt}
\end{figure}

\begin{figure}[]
	\centering
	\includegraphics[width=.98\columnwidth]{{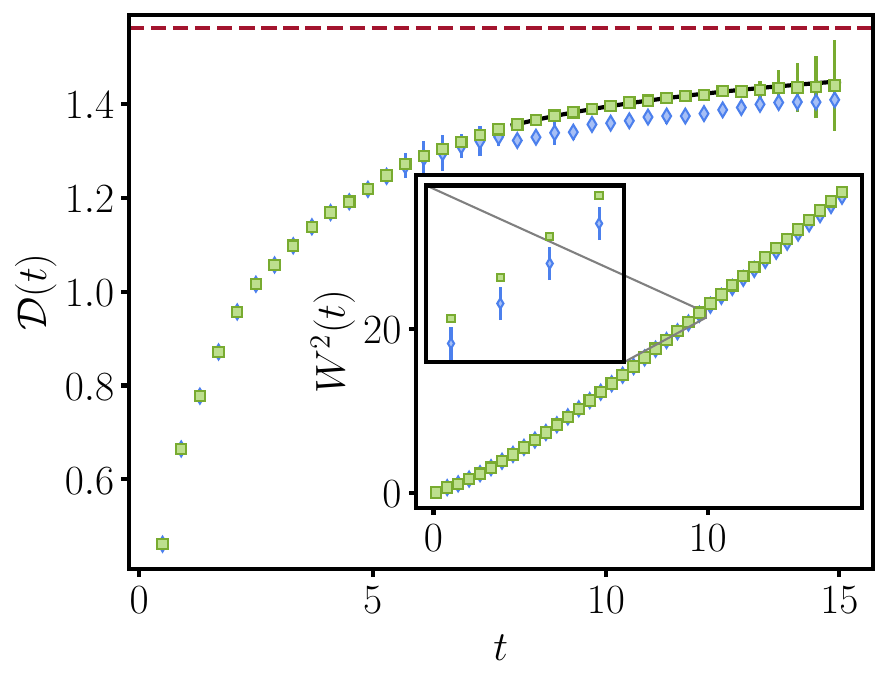}}
	\caption{Spatial variance (Eq. 3 of the main text) of the normalized autocorrelations (inset) and corresponding diffusion constant (main plot) in the non-integrable Ising case with $(J = 1, g = 1.4, h = 0.9)$ obtained from the TLCC (green squares)  and TEBD (blue diamonds). The solid black line in the main plot shows an heuristic fit of the form $\mathcal{D}_E \exp(b/t)$, which predicts the asymptotic value $\mathcal{D}_E\approx1.6$ (red dotted line). The bond dimensions used for both methods are 1024, and the errorbars show the difference with the results of bond dimension 512.}
	\label{figS:diffusion}
\end{figure}

A standard way to bound the error in TN simulations is to sum the squares of the discarded Schmidt weights in every truncation \cite{Paeckel2019tevolRev}. For the results from TEBD and Heisenberg-picture DMRG, we use that as a measure of the growth in the errors during the simulation. Notice that in the case of TEBD, as we are simulating an infinite system that measure is only an heuristic, as it does not provide an upper bound in the error that one can
incur when evaluating some observable. For the TLCC, both when we use the standard and the hybrid truncation from \cite{Hastings2015impro} to truncate the boundary vectors, we keep track of the deviation of the expectation value of the identity. As explained in the main text, knowledge of the boundary vectors $(L(t)|$ and $|R(t))$ gives access to out-of-equilibrium expectation values by computing the expectation value $(L(t)|E_O(t)|R(t))$. That includes the possibility of computing the expectation value of the identity, which should be one for any properly normalized state. 
The deviation with respect to this value, when using MPS approximations for $(L(t)|$ and $|R(t))$, gives a good measure of the error of the TLCC.
%The deviation from one when $(L(t)|$ and $|R(t))$ are approximated by MPS with fixed bond dimension turns out to be  We use this and use them to compute the expectation value of the identity is a good measure of the accuracy of TLCC. 

In order to perform the scaling analysis, we compute as a function of time the quantities mentioned above for the different TN algorithms with simulations with different bond dimensions. Setting a constant value of the precision, that is, of the truncation errors in the TEBD and Heisenberg-picture DMRG cases and of the deviation from the identity for the TLCC, we can keep track of the times when the different bond dimensions exceed the desired precision threshold.

\section{Alternative values of the parameters}

In the main text, for simplicity, we focused in a particular point in parameter space in the non-integrable case. Here, we show results obtained in a different case, described by the couplings $(J = 1, g = 1.4, h = 0.9)$ and studied in \cite{Kim2013ballistic, Rakovszky2020dissip}. The spatial variance and diffusion constant obtained with TLCC for this case are shown in Fig. \ref{figS:diffusion}, along with results obtained with TEBD. As seen in the main text, the introduction of TLCC allows in this case as well to extend the range where it is possible to reliably simulate out-of-equilibrium dynamics by a factor around two. 

\end{document}